%% file: article.tex
\documentclass{elsarticle}

\usepackage[utf8]{inputenc}
\usepackage{hyperref}

\usepackage{graphicx}

\journal{Future Generation Computer Systems}

\usepackage{amsmath,amssymb}
\usepackage{mymath}
\usepackage[draft]{fixme}
\usepackage{paralist}
\usepackage{tabularx}
\usepackage{multirow}
\usepackage{listings}
\lstdefinelanguage{Scala}{
  morekeywords={abstract,case,catch,class,def,%
    do,else,extends,false,final,finally,%
    for,if,implicit,import,match,mixin,%
    new,null,object,override,package,%
    private,protected,requires,return,sealed,%
    super,this,throw,trait,true,try,%
    type,val,var,while,with,yield},
  otherkeywords={=>,<-,<\%,<:,>:,\#,@},
  sensitive=true,
  morecomment=[l]{//},
  morecomment=[n]{/*}{*/},
  morestring=[b]",
  morestring=[b]',
  morestring=[b]"""
}
\begin{document}

NOTICE: this is the author’s version of a work that was accepted for publication by Elsevier. Changes resulting from the publishing process, such as peer review, editing, corrections, structural formatting, and other quality control mechanisms may not be reflected in this document. Changes may have been made to this work since it was submitted for publication. A definitive version was subsequently published in Future Generation Computer Systems, Volume 37, July 2014, Pages 390–400, http://dx.doi.org/10.1016/j.future.2014.02.002 .

\begin{frontmatter}

\title{Computing Agents for Decision Support Systems}

\author[agh]{D. Krzywicki}
\ead{daniel.krzywicki@agh.edu.pl}

\author[agh]{\L{}. Faber}
\ead{faber@agh.edu.pl}

\author[agh]{A. Byrski\corref{cor}}
\ead{olekb@agh.edu.pl}
\cortext[cor]{Corresponding author}

\author[agh]{M. Kisiel-Dorohinicki}
\ead{doroh@agh.edu.pl}

\address[agh]{AGH University of Science and Technology, Al.\ Mickiewicza
30, 30-059 Krakow, Poland}

\begin{abstract}
In decision support systems, it is essential to get a candidate solution
fast, even if it means resorting to an approximation. This constraint
introduces a scalability requirement with regard to the kind of
heuristics which can be used in such systems. As execution time is bounded,
these algorithms need to give better results and scale up with additional
computing resources instead of additional time. In this paper, we show how
multi-agent systems can fulfil these requirements. We recall as an example the
concept of Evolutionary Multi-Agent Systems, which combine evolutionary and agent computing
paradigms. We describe several possible implementations and
present experimental results demonstrating how additional resources
improve the efficacy of such systems.
\end{abstract}

\end{frontmatter}

\input{intro.tex}

\input{dss.tex}

\input{agent.tex}

\input{emas.tex}

\input{execution-models}

\input{state}

\input{implementation}

\input{experiments}

\section{Conclusion}

Metaheuristics can be valuable in decision support systems with time
constraints. We discussed in this paper how the agent approach can be applied to
these systems in order to build efficient and scalable software. We described
the concept of evolutionary multi-agent systems, an example of a metaheuristic
combining agent-based and evolutionary techniques.

The main goal of our work was to investigate the existing methods of building
agent software and suggest new directions of development. In particular, we
wanted to see if the dominant approach, which considers every agent as a unit of
concurrency, is really efficient in computational intensive simulations.
 
To that purpose, we developed two alternate implementations of an evolutionary
multi-agent system which generalise the trends in existing agent software. We
introduced the idea of meeting arenas, an agent-based realisation of the
Mediator design pattern which allow to efficiently structure multi-agent
systems. We applied this concept to two versions of the algorithm: an
asynchronous one, where every agent is a fully independent entity, and a
synchronous one which treats agents as simple data structures.
 
Our experiments revealed that an asynchronous implementation, which may feel
more \emph{the agent way}, is nearly an order of magnitude less efficient than a
synchronous one based on the same design. Several prototypes in other
technologies supported these results. Further experiments on the synchronous
implementation demonstrated that it can easily be scaled in a distributed
setting, so that the efficiency of the algorithm increases when new nodes are
added to the computation.

Therefore, we showed that the prevailing approach in existing agent platforms is
not best suited in this particular class of applications. Instead, there is
still room for improvement in the field of agent software dedicated to intensive
simulations and computations. To that purpose, the concept of meeting arenas
introduced in this paper allow to retain the expressive power of existing
agent-based algorithms but can lead to much more efficient synchronous
implementations.

In the nearest future, we want to see if concepts used in the functional programming paradigm could be more
suited or more efficient in multi-agent software than the dominant
object-oriented approach. Future work could also tell what kind of parallelism
could be efficiently introduced in populations of agents. In particular, it
would also be interesting to see how the Map/Reduce paradigm could be used to
develop efficient massive multi-agent systems with hundreds of thousands of
agents.

\bibliographystyle{model1-num-names}
\bibliography{exported}

\section*{\textbf{ABOUT THE AUTHORS}}
Daniel Krzywicki obtained his M.Sc.\ in 2012 at AGH University of Science and
Technology in Cracow and is currently a Ph.D.\ student at the Department of
Computer Science of AGH-UST. His research interests include agent-based
computations, functional programming and distributed systems.

\L{}ukasz Faber obtained his M.Sc.\ in 2012 at AGH University of Science and
Technology in Cracow and is currently a Ph.D.\ student at the Department of
Computer Science of AGH-UST. His research interests include agent-based
modelling and distributed systems.

Aleksander Byrski obtained his Ph.D.\ in 2007 and D.Sc.\ (habilitation) in 2013 at AGH University of Science and
Technology in Cracow. He works as an assistant professor at the Department of
Computer Science of AGH-UST. His research focuses on multi-agent systems,
biologically-inspired computing and other soft computing methods.

Marek Kisiel-Dorohinicki obtained his Ph.D.\ in 2001 
and D.Sc.\ (habilitation) in 2013 at AGH University of Science
and Technology in Cracow. He works as an assistant professor at the Department
of Computer Science of AGH-UST. His research focuses on intelligent software
systems, particularly using agent technology and evolutionary algorithms, but
also other soft computing techniques.

\section*{\textbf{ACKNOWLEDGEMENT}}
The research presented in the paper was partially supported by the European Commission
FP7 through the project ParaPhrase: Parallel Patterns for
Adaptive Heterogeneous Multicore Systems, under contract
no.: 288570 (http://paraphrase-ict.eu).

The research presented in the paper was conducted using PL-Grid Infrastructure (http://www.plgrid.pl/en).

\end{document}

%% file: intro.tex
\section{Introduction}
The need to gather and analyse vast amounts of information from numerous sources
has grown in importance. Such data is often a basis for simulations and
computations that support decision making. It may be needed to run many
computing tasks, in order either to test different parameters in a model or to
verify a statistical hypothesis. An exhaustive search for optimal solutions to a
decision making problem is usually time-consuming and thus not acceptable in
real-time conditions. Instead, metaheuristics may quickly provide good-enough
options to be further considered in the decision making process
\cite{Michalewicz2004}.

Examples of use cases with real-time constrains, where a quick approximated
solution may be better than an outdated optimal one, may include:
\begin{itemize}
  \item Portfolio optimisation --- a decision support system can apply different
  models to the available market data and allow the user to quickly react to
  arising trends \cite{drezewskisepielaksiwik}.
  \item Crisis management --- in crisis situations, such as fire outbreaks,
  flooding or earthquakes, intensive simulations are required in order to
  suggest possible evacuation routes or to assign rescue units to tasks. 
  Geographical information usually needs to be considered, yielding optimisation
  problems similar to transportation-related ones \cite{crisis1}.
  \item Production planning --- decision support systems can help in scheduling
  work, rescheduling production plans in case of hardware failures, implementing
  just in time strategies or balancing conflicting goals (e.g.
  high system throughput vs low machinery usage) \cite{lawrynowicz}.
\end{itemize}

Metaheuristics may still require a significant computational power if the
acceptable solution is to be found in a reasonable time. For this purpose,
large-scale infrastructure is usually used, such as clusters, grids or clouds.
To fully benefit from this computational power, it is required to appropriately
plan their development and deployment, along with adequate tools and careful 
testing.

Because of its intrinsic decentralisation \cite{wooldridge95intelligent}, the
agent approach is well suited to design scalable distributed models and has been
applied in various decision support systems.
This approach may be summarised as the introduction of artificial intelligence
techniques into the system, transforming it from a passive tool into an active
collaborator in decision making. A number of such case-oriented systems have
been proposed and verified in practice \cite{97,114}.

Well-known general-purpose agent-based development tools (such as JADE \cite{Bellifemine2001},
RePast \cite{North2013} or Madkit \cite{Gutknecht2001}) may
not be the best choice to implement such computational intensive simulations,
when throughput and scalability are more important than code migrations or
FIPA-compliant communication. Therefore, over the last 10 years, we have been
involved in the development of several alternative platforms dedicated to large
scale agent-based simulations and
computations \cite{my:Holo2009,augmented_cloud,faber}.

In this work, we discuss the implementation aspects of using computing agents in
large-scale environments, with a focus on performance. We compare different
approaches to agent execution and parallelism, based on the metaheuristic
called evolutionary multi-agent systems (EMAS), which is a hybrid of
agent-oriented and evolutionary-based computing \cite{ker}. We introduce the
concept of \emph{meeting arenas}, which allow to design more efficient and scalable
multi-agent systems. Nevertheless, we show that explicit parallelism, when each agent is
mapped onto a thread, can be much less effective than a simple but optimised
sequential implementation. Finally, we show that such agent-based metaheuristics
can be easily scaled with additional computational resources.

We start the paper with a discussion on the applicability of the agent-oriented
paradigm and metaheuristics in decision support systems (Section~\ref{sec:dss}),
along with an EMAS example. In Section~\ref{sec:models}, we introduce the most
common approaches to parallelism in agent-oriented computing and follow with a
review of popular agent platforms in Section~\ref{sec:platforms}. We describe in
Section~\ref{sec:implementation} how to implement an evolutionary multi-agent
system using two different approaches---a synchronous and asynchronous one.
Finally, we conclude the paper by comparing the performance and scalability of
both approaches in Section~\ref{sec:experiments}.

%% file: dss.tex
\section{Agent-Based Metaheuristics in Decision Support}
\label{sec:dss}

Decision Support Systems (DSS) are information systems that support different
business or organisational activities involving decision-making. They are
especially useful in situations where quickly changing, hard to specify in
advance conditions are encountered.
Referring to Power's taxonomy for DSSs \cite{power} this paper focuses
on Model-driven DSSs, which help the users in the analysis of the current
situation by allowing to manipulate statistical, simulational or optimisational
models.

\subsection{Metaheuristics for DSSs}

The models used in DSSs are usually very complex and computationally hard,
because the underlying problems are very difficult as well.  In such cases, one
often turns to solutions based on so-called heuristic methods, which provide
``good-enough'' solutions without caring whether they may be proved to be
correct or optimal \cite{Michalewicz2004}.  These methods trade-off precision,
quality and accuracy in favour of smaller execution time and computational
effort. They are necessary to deal with difficult problems, and are referred to
as methods of the last resort \cite{lastresort}.

A general definition of a heuristic algorithm, which does not specify details
such as a particular problem, search space or operators, is called a
\emph{metaheuristic}. For example, a simple algorithm such as greedy search may
be defined without going into more details as ``an iterative, local improvements
of a solution based on random sampling''\cite{Glover2003}.

A simple but adequate classification of metaheuristics (cf.\
\cite{dreo2005metaheuristics}) distinguishes two groups of techniques.
Single-solution metaheuristics work on a single solution to a problem, seeking
to improve it. The examples are greedy search, tabu search or simulated
annealing.  Population-based metaheuristics explicitly work with a
population of solutions and put them together in order to generate new
solutions.  The examples are evolutionary algorithms,
immunological algorithms, particle swarm optimisation, ant colony optimisation, memetic algorithms
and other similar techniques.  They are usually inspired by
nature and imitate different phenomena observed in e.g., biology, sociology,
culture or physics \cite{talbi1}.

%% file: agent.tex
\subsection{Agent Approach}

The key concept in multi-agent systems (MAS) consist in intelligent
interactions, such as coordination, cooperation, or negotiation.  Therefore,
multi-agent systems are ideal in representing problems which can be solved using multiple methods by numerous entities with various perspectives. One of the
most important features in a multi-agent system is the autonomy of the agents,
as they can fulfil the tasks assigned to them according to their own strategy
and the situation observed in their environment. In consequence, agents are
adaptable and proactive \cite{wooldridge95intelligent}.

Combining the agent-oriented approach with population-based metaheuristics seems
natural but has yet been the topic of little work. The entities processed in the
course of the computation can often be considered autonomous and treated as
agents in a common environment. The operations involving many such entities can
be defined as interactions between these agents. 

This change of modelling perspective allows to perceive parts of the system on a
higher abstraction level and build hybrid systems which combine techniques from
different metaheuristics. New problems often arise, such as the lack of global
knowledge or the need for proper synchronisation of the agents' actions.
However, interesting and useful effects also often result from the cooperation
of different mechanisms in one system \cite{icnnsc2002}.

In this paper, we focus on an example of such a hybrid approach, in which agents
are subject to an evolutionary process. Such a combination yields interesting
new features when compared to classical evolutionary algorithms, such as a
decentralised an emergent selective pressure.

%% file: emas.tex
\subsection{Evolutionary Multi-Agent Systems}
Agents in an evolutionary multi-agent system (EMAS) represent solutions to a
given optimisation problem.

Inheritance is achieved through reproduction, with the possible use of
variation operators such as mutation and recombination, like in classical
evolutionary algorithms.  Yet agents are to be autonomous in their
decisions and no global knowledge is available to them. Therefore, in
contrast to classical evolutionary algorithms, selection needs to be
decentralised and involve peer-to-peer interactions instead of being
system-wide. 

In order to do that, a solution based on the acquisition and exchange of
non-renewable resources has been proposed in \cite{kc}. The quality of the solution represented by the agent is expressed by the amount of resources the agent owns.
In general, these resources should pass from worse agents to better ones.  This
might be realised through encounters between agents, which cause better ones to
end up with more resources and make them more likely to reproduce. Worse agents lose
resources which increases the probability of their death.
Because of such indirect dynamics of reproduction and death, agents' lifespans
overlap and so do the generations.  Moreover, the size of the population is
dynamic and can be changed by varying the amount of available resources. A
detailed study of computing with EMAS, in particular the influence of its
different parameters on the computing efficiency may be found in \cite{tuningemas}.

\begin{figure}[h] \centering
\includegraphics[width=.6\columnwidth]{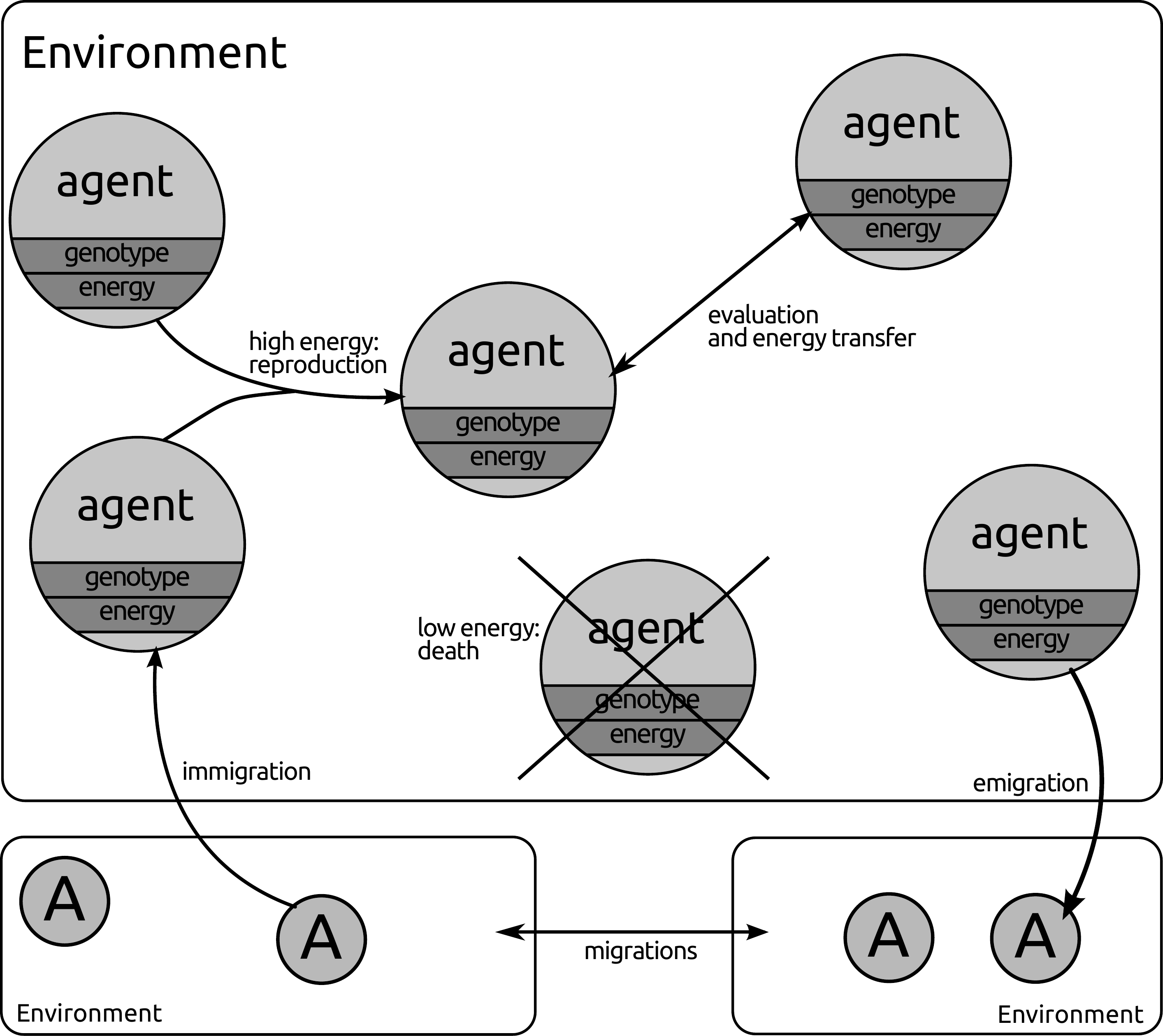} \caption{An example of
a simple evolutionary multi-agent system (EMAS). Agents with higher fitness take
energy from agents with lower fitness. High levels of energy increase the
probability of reproduction and reduce chances of death.
In consequence, the selection process is decentralised and selective pressure
softens. Multiple agent environments can be connected through
agent migrations, like in the classical island model.} \label{fig:emasls}%
\end{figure}

Agents are grouped within \emph{environments} which define the information and
resources an agent has access to. Agents can interact with
each other directly only within the same environment. However, they are able to move to
another environment, thus exchanging information and
resources all over the system \cite{ker} (see Fig.~\ref{fig:emasls}).

Environments are largely independent and communicate only through agent
migrations. Therefore, they can be easily treated as basic units of
distribution, as in the classical island model in evolutionary algorithms. 
In addition to improving the performance of the algorithm, it also 
increases the diversity of solutions in the whole population (allopatric
speciation).  Other metaheuristics can also be introduced, such as
immunological selection \cite{my:IIPWM2005} and niching \cite{csci109}.

The principle of an evolutionary multi-agent system consist in the explicit
hybridisation of agent-oriented and evolutionary computing. This contrasts
with usual agent-oriented approaches, which use the agent-paradigm to solve
certain tasks by delegating them to particular agents and combining the outcomes
of their work (see, e.g.\ \cite{danoy2010multi}).

%% file: execution-models.tex
\section{Interaction and Execution Models for Agents}
\label{sec:models}

In agent-oriented computing systems, agent interactions are one of the crucial
aspects of their work. It is easy to predict that parallelising them
can significantly increase the throughput of the system. However, this comes at
the cost of increased communication and synchronisation. Therefore, an important
issue is to choose the appropriate granularity of the entities in the
computation.

As agents are defined as autonomous and independent beings, it seems natural to
look for further concurrency within a single environment. The question is where
to put the boundaries of concurrent execution, as it has consequences on both
performance and ease of programming. This section discusses the most common
models of execution and interaction in existing agent software.

\subsection{Heavyweight Agents}

In this model every agent is associated with a thread and communicates through
message passing. Some agents may passively wait for incoming messages and react
to them. Other agents may actively initiate interactions with other agents.
It is difficult to achieve a coordinated life cycle among such agents, since the
corresponding threads may be arbitrary interleaved. 
Therefore, some kind of synchronisation between agents still needs
to be introduced, usually in terms of a specific communication protocol.

In order to interact with each other, agents need to locate other agents willing
to perform the same actions. For example, in an evolutionary multi-agent system,
an agent with enough resources to reproduce needs to find another one which also
has enough resources. In order to do that, it could ask all other agents in the
population. However, such a solution is obviously inefficient, because of the
intensity and redundancy of the required communication.

\begin{figure}[h]
\centering
\includegraphics[width=.7\columnwidth]{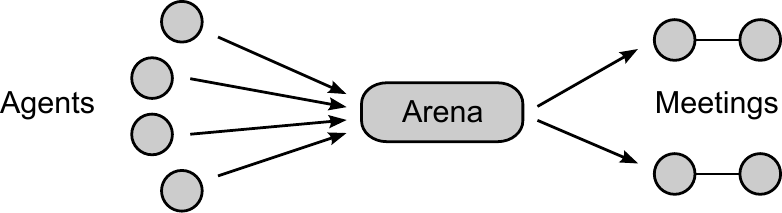}%
\caption{Meeting arenas allow to group similar agents and coordinate meetings
between them.}%
\label{fig:meetingArena}%
\end{figure}

\label{sec:meetingArenas}
A better approach, introduced in this paper, is to use a mediating entity, which
we call a \emph{meeting arena}. Every time an agent wants to perform an action,
it chooses an appropriate arena to meet with other similar agents. The arena is
then able to partition its members in groups of some given arity and mediate the
meeting itself (see Fig.~\ref{fig:meetingArena}). Examples with pseudocode are
given in Section~\ref{sec:implementation}.

The usage of meeting arenas should bring many benefits, not only in terms of
efficiency, as the algorithm itself can be structured more clearly. Agents only
need to be given a set of rules, in order to choose an arena on the basis of
their state. The actual protocol of agents interactions can then be defined at
the level of the appropriate arena.

Assigning a thread to each agent may feel very natural. In practice, however,
the number of agents is often much higher than the number of cores, especially
in simulations. Performance may then be seriously hindered by frequent context
switches, although this overhead may be reduced by sharing a pool of threads
among agents. However, this model still involves intensive communication and
costly processor cache synchronisation. In consequence, the trade-off for such
concurrency may be higher than expected.

\subsection{Lightweight Agents}
\label{sec:thread-pool}

An opposite approach is to consider agents as parts of the model, not parts of
the implementation. As such, they are simply represented as data structures and
processed like in a discrete event simulation.

The execution of an individual agent has to be divided into smaller parts which
can be interleaved. These parts, which we will call \emph{actions}, could for
example consist in executing a single step or querying a neighbour. Given its
current state, every agent decides which action to perform next. This action is
submitted for later execution to an executor service owning a pool of threads,
like in the Command Object design pattern.

The difference with regard to classical discrete event simulation is that
actions are generated synchronously but can be executed asynchronously. In other
words, the state of an agent during the execution of the action may be different
from the time when the action was created.

The performance of such a model will usually be higher than in the
previous one, more consistent memory access patterns resulting in more efficient
processor usage. Even though the explicit parallelism is reduced, throughput can
be improved, because frequent agent interactions no longer need to be
synchronised between threads.

Moreover, independent actions can still be executed in parallel by the executor
service. This is consistent with the meeting arena concept described above, as
actions on common subsets of agents may be grouped together and considered as a
single meeting.

%% file: state.tex
\section{Distributed and Parallel Multi-Agent Frameworks}
\label{sec:platforms}

This section provides an overview of existing multi-agent frameworks. In our review, 
we focus on parallelisation and distribution capabilities, with regard to the
aspects discussed in the previous section.

First, we briefly describe some selected tools which specialise in metaheuristics.
They are interesting examples of improving metaheuristics
with agent systems, but lack more general agent-oriented features. 
However, all of them share the idea of an agent being an
\emph{executor} of the algorithm and not a \emph{participant}.

The platforms described in the consecutive sections provide more sophisticated
support for agent-based systems but are not necessarily well suited to
metaheuristic computations.
Some properties are shared by almost all of them, such as: the choice of an
object-oriented programming language (mostly Java) and a representation of
agents as objects with an internalised state. Other characteristics include:
models that are too heavy (e.g.,\ JADE due to FIPA compatibility), a large and
complicated code base due to the implementation of many communication,
distribution and component-oriented mechanisms in the platform instead of using
ready solutions (e.g.,\ Jadex). 
                                           
\subsection{Metaheuristics Frameworks}
\label{sec:meta-frameworks}

The four frameworks presented below are examples of introducing multi-agent
systems to metaheuristic computations. They are specialised to this purpose
and lack well-defined distribution facilities. We present them as alternative
models but we do not discuss their implementations in-depth.

\paragraph{MAGMA (Multi-AGent Architecture for Metaheuristics)}
MAGMA \cite{Milano_Roli_2004} is a multi-level (hierarchical) architecture of a
multi-agent system. Each level of agents has different objectives and represents
a different level of abstraction of the algorithm. For example, level 0 agents generate a sample
solution and then level 1 agents improve it by searching the neighbourhood of
that solution. There may be several agents that participate in an algorithm on
each level. Composing different metaheuristics is also possible with a
coordination provided by a higher level of the architecture (level 3). This way
agents wrap selected functions of the algorithm and not the whole algorithm
itself (as it will be the case for further platforms).

\paragraph{MAS-DGA (Multi-Agent System for Distributed Genetic Algorithms)}
The attention of the authors in \cite{Noda2002} is focused on approaches to the
question of migration. They propose the MAS-DGA framework that comes from the
concept of Distributed Genetic Algorithms, where the population is divided into
interacting subpopulations handled by different genetic algorithms (GA).
In the case of MAS-SGA, these GA are encapsulated in agents. The authors suggest
a possibility of distribution on the agent level but they do not provide
descriptions of any specific examples nor implementations of this model.

\paragraph{AMF (Agent Metaheuristic Framework)}
The authors of AMF \cite{Meignan2008} extend metaheuristics with an
agent-oriented approach and an organisational model based on \emph{roles} and
\emph{interactions}. The RIO meta-model described in the paper involves the
three following concepts: Role, Interaction and Organisation. Metaheuristics are
organisations, and agents play specific roles in these organisations. Some of
the defined roles are: the intensifier (performs a search in a search space),
the diversifier (identifies new promising regions in the search space), the
guide (structures the information from two previous roles), etc.

\paragraph{MAS4EVO (Multi-Agent System for EVolutionary Optimization)}
In \cite{danoy2010multi} the authors propose a model and a framework (DAFO --
Distributed Agent Framework for Optimization) that is a significant improvement
over the previous three. The framework is built on MadKit (see
Section~\ref{sec:madkit}). In this model, authors introduce three types of
agents: problem solving agents which optimise functions, fabric agents which are
responsible for initialising and configuring the computation, and observing
agents which generate output for the end-user.

\subsection{Jadex}

Jadex\footnote{\url{http://www.activecomponents.org}} is an agent-based
programming framework that exploits a novel approach to
agents-components unification called ``active components''
\cite{Braubach2012}.

The concept of Active components unifies SCA (Service Component Architecture)
components with agents. This results in components that are able to use, in
addition to traditional required and provided service interfaces, asynchronous
messaging and that can act autonomously. It has a tremendous impact on
behaviours of these components. They, for example, can refuse service call
execution when they cannot or do not want to process the request.

\paragraph{Agents}

Jadex offers two ways to implement agents. It is possible to use full-featured,
BDI (belief-desire-intention) agents and simple, so-called \emph{micro agents}.
Micro agents are usually just annotated POJO (Plain Old Java Objects) classes. They follow three-phased
execution semantics: initialisation, execution and termination. Additionally, an
agent can schedule actions to be run later. As an agent is also an active
component, it may receive service calls and incoming messages.

\paragraph{Distribution}

Distribution in Jadex is provided transparently to the developer and it is
implemented using a layered architecture. Services, for example, may use remote
asynchronous method calls. Transparency is achieved by using proxy interfaces
implementation. Internally, remote calls are implemented using asynchronous
messaging between remote management system components. Messages are encoded and
transmitted through some chosen \emph{stream}. The encoding of messages
is provided by \emph{codecs} which need to (un)marshall Java objects to binary
or XML format but which can also provide more sophisticated features:
e.g., compression or encryption. A stream can use different communication
transport: TCP, HTTP and others.

The second aspect of distribution is peer awareness and discovery. Jadex takes
care of it automatically on all levels, including service (i.e.\ interface)
binding. When there is a look-up for a required service, proxy components on a
local node redirect search requests to the remote management system to perform
remote look-ups of services.

Before that, remote platforms in the network need to be discovered. For this
purpose, Jadex provides a few different mechanisms: e.g., broadcast discovery
which sends UDP announcements about a platform on a local network or registry
discovery in which there is one, central registry created for all platforms to
announce themselves.

Other features of Jadex include, among other things, support for interaction
with external systems using web services and a GUI-based control centre.

\subsection{MadKit}
\label{sec:madkit}

MadKit\footnote{\url{http://www.madkit.net}} is a generic, customisable
multi-agent platform based on a specific organisational model \cite{Gutknecht2001}.
Agents are divided into groups and they may have particular roles in them.
The centralisation of the platform around the organisational concepts
is, in the view of the authors of the platform, a key element for building
heterogeneous systems. 

The MadKit architecture is built on the agent-group-role (AGR) model. This model is
used to built organisations: an \emph{organisation} is described using terms of
interacting \emph{groups} and \emph{roles} and is separated from the concept of an agent.

\paragraph{Agents}

A MadKit agent is an entity that can communicate and which has several roles within
one or more groups. Groups are atomic structures aggregating agents and they can
overlap. Roles are tags for agent functions within groups. Agents request them
on their own and they may be granted or denied them. Communication between agents is
achieved using asynchronous messaging. Addressing of agents is done using their
addresses or by their specific roles in one of their groups.

\paragraph{Architecture and Distribution}

The architecture of MadKit is developed around the AGR concept. Moreover, it
follows some additional design decisions, the most interesting being:
micro-kernel architecture and agentified services.

The micro-kernel, responsible for basic platform management,
handles only most essential functions. The rest of the needed
services is handled by agents. The micro-kernel tasks are: control of
groups and roles, life-cycle management of agents, local messaging. It also
supports so-called ``kernel hooks'' which allow extension of its functionality by
operations executed in the publish-subscribe model. Two types of hooks are
supported: monitor and interceptor hooks. The former can be used by many agents
at the same time whilst the latter can be hold on by only one agent and can be used
to prevent the operation from successful execution. Additionally, it is possible
to execute actions on kernel when an agent is a member of the \emph{system} group.

The agentification of services describes a concept of turning system services
(e.g., distributed message passing, migration) into agents. This makes the
platform very extensible and flexible as every component can be easily
replaced: communication with services is no different than with other agents.

MadKit has support for transparent distribution. Groups can span across many
platform nodes. It is provided by two roles in the \emph{system} group:
communicator (routes messages to other nodes) and synchroniser (keeps information about
groups memberships synchronised across all nodes).

MadKit provides also a graphical environment for visualising simulations and
controlling the platform.

\subsection{\(\mathrm{\mu^2}\)}

\(\mathrm{\mu^2}\)
(micro-squared)\footnote{\url{http://sourceforge.net/apps/mediawiki/micro-agents/}}
is a multi-agent platform centred around the concept of a \(\mu\)-agent
(or micro-agent):
a small-size agent, that can be recursively constructed from
other micro-agents with decomposition and fine-grained separation of concerns
in mind \cite{Frantz2011}.
\(\mathrm{\mu^2}\) is implemented in Java and in Clojure and available under
GPL 3.0 license.

\paragraph{Agents}

Micro-agents are autonomous, persistent, reactive and proactive. They
can play one or more roles which fulfil so-called applicable intents. Intent is
another name for an intention or ``abstract request specification''.  The
platform provides some organisational modelling approaches. An agent can be in
a group leader role. In such case, it controls many sub-agents, propagates control
messages and structures the society. Due to this approach agents can construct
themselves using sub-agents, and sub-agents also can be group leaders. This is
how decomposition can be implemented in the platform. Other roles include:
social roles that allows agents to communicate asynchronously and passive roles
which support only synchronous communication. The latter role was introduced to
reduce possible performance penalties resulting from asynchronous messaging.

\paragraph{Distribution}

As it is mentioned in \cite{Frantz2011}, \(\mathrm{\mu^2}\) is a platform
that can be run on Android devices. Micro-agents are encapsulated into
Android service and they are integrated into the rest of the system.
Communication between normal applications and agents is transparent. 

\subsection{JADE}

JADE\footnote{\url{http://jade.tilab.com/}} is a mature (founded in 2000)
Java framework for developing agent-based applications with a very strong
relationship with FIPA specifications \cite{Bellifemine2001}.
Its architecture is focused on
a peer-to-peer communication with some centralised services. Two software
components are specified: agents (autonomous, using asynchronous messaging)
and services (non-autonomous, running on a single or multiple nodes).

\paragraph{Agents}

JADE agents exist in containers (basically Java processes)
which can be distributed over the network. The way messages are structured is
compliant with FIPA Agent Communication Language.

\paragraph{Distribution}

The peer-to-peer nature of JADE made it possible to create many
reimplementations of nodes, e.g., for mobile environments like Android
\cite{Ughetti2008}. Distribution is gained by splitting a JADE container into a
frontend and backend. The former runs on a mobile device and is rather
lightweight, the latter usually runs on a more powerful computer.

\subsection{Repast Suite Family}

Repast\footnote{\url{http://repast.sourceforge.net}} is an open-source,
agent-based modelling and simulation toolkit \cite{North2013}.
It has many versions for various programming languages. The most interesting
ones are the newest: Repast Simphony (for Java) and Repast HPC (for C++).
All of them uses the ``new BSD'' license.

Repast Simphony is a complete rewrite of older Repast 3 with a modular
architecture, extendable via plugins. Individual components (e.g., networking,
logging) can be replaced easily. Plugins are layered and separate layers can be
replaced with similar easiness. There is a separation between model
specification, execution, data storage and visualisation.

A core of the Repast Simphony consists of components responsible for
simulation functions (e.g.,\ time scheduling, space management, random number
generators). 
ss
\paragraph{Agents}

Agents are modelled as objects, collections of agents -- as contexts, and the
environment -- as projections. A \emph{context} is a set of objects and may
represent an agents' population but does not describe any structure or
relationships between agents. The second term -- \emph{projection} -- was created to
define structures of agents in contexts. They may be, for example, network or
grid structures.

\paragraph{Distribution}

Repast does not offer distribution facilities similar to other platforms.
However, a user can prepare its own distributed environment using external
facilities, for example, Java RMI (Remote Method Invocation), which is an
object-oriented remote procedure call mechanism.

%% file: implementation.tex
\section{Implementation Aspects}
\label{sec:implementation}

Evolutionary multi-agent systems and similar agent-based computing
systems need
lightweight, reusable and easy-to-parallelise solutions. In particular, the
implicit agent-orientation perceived at the implementation level of these
platform does not seem inevitable to us. We think that agent features should be
a part of the conceptual level, but do not need to be reflected in the
implementation in the case of computing systems.

Considering the execution models described in section 3 and
their implementation in existing software tools for multi-agent
systems, we wanted 
to compare these two approaches to tell what granularity is best
suited for agent-based computing and simulation. In order to abstract from the
properties of these frameworks not relevant to the problem, we implemented two
custom versions of an evolutionary multi-agent system.

In the first version agents are asynchronous and can be mapped to separate
threads or share a thread pool. The second version is synchronous and optimised
for single-thread execution. Both versions are written in the Scala programming
language, a relatively new programming language for the Java Virtual Machine.
Scala\footnote{\url{http://www.scala-lang.org/}} is suited for both
object-oriented and functional programming, supports parallel and asynchronous
programming and is compatible with Java code and existing libraries.

Both versions are based on the concept of meeting arenas introduced in
Section~\ref{sec:meetingArenas}. Every agent is assigned with a solution to the
optimisation problem, its fitness and some ``life energy'' (a single resource).
The behaviour of the agents is the same in both versions (see
Listing~\ref{ls:agentBehavior}). They differ in how agents join arenas and how
arenas execute meetings.

\begin{figure}[h!]
\begin{lstlisting}[language=Scala, label=ls:agentBehavior, 
caption={Agents choose an arena to join based on their current resources, in
this case energy.}] 
def chooseArena = energy match {
	case 0 => deathArena
	case e if e > threshold	
	       => reproductionArena
	case e => fightingArena
}
\end{lstlisting}
\end{figure}

In this evolutionary multi-agent system, we use the following arenas:
\begin{compactitem}
	\item agents are removed from the system in the \textbf{death arena}
	\item agents compare their fitness in the \textbf{fighting arena}. Losers give
	some of their energy to the winners
	\item new agents are created in the \textbf{reproduction arena}. Children
	solutions are derived from their parents using variation operators. Parents give
	some of their energy to their children.
\end{compactitem}

\subsection{Asynchronous EMAS}
\label{sec:asyncEmas}

This version is similar to the approach in frameworks like Jade, in which agents
are the basic unit of concurrency. They are independent entities which do not
directly expose state and can only query each other for information.

Agents and arenas have been implemented using the Akka\footnote{\url{http://akka.io/}}
actor library. They are represented by actors which execute asynchronously and
communicate through message passing. As such, agents can be mapped to threads in
a very flexible way. Akka actors are handled by a component called the
\emph{dispatcher}. The dispatcher allows each actor in turn to process one or
more messages from its mailbox. It is also used to execute asynchronous tasks.

The processing of a message or task can happen in any thread owned by the
dispatcher, which behaviour is fully configurable. The dispatcher can use a
single thread, a pool of threads or assign a separate thread to each actor. Akka
ensures happens-before relationships between the processing of consecutive
messages and preserves memory consistency.

After its previous meeting have ended, every agent chooses an arena and join it
by sending a \texttt{JoinMeeting} message.  Every arena has a fixed size and
acts as a cyclic barrier: a meeting is triggered as soon as the capacity of the
arena have been reached (see Listing~\ref{ls:asyncArena}). Multiple meetings
may be happening at the same time, but every agent can only take part in one of
them. When the meeting is finished, a \texttt{MeetingEnded} message is sent
asynchronously to its participants so that they can choose a new arena to join.

\begin{figure}[h!]
\begin{lstlisting}[language=Scala, label=ls:asyncArena, 
caption={Asynchronous arenas act as a cyclic barrier and trigger an asynchronous
meeting as soon as they are full.}] 
def receive = {
	case JoinMeeting =>
		waitingRoom.add(sender)
		if(waitingRoom.isFull()) {
			val members = waitingRoom.flush()
			performMeeting(members) andThen {
				members foreach { 
					member => member tell MeetingEnded 
				}
			}		
		}
	}
\end{lstlisting}
\end{figure}

An additional mechanism, omitted above for clarity, triggers a meeting after
some inactivity timeout. This may be beneficial for the algorithm (e.g.,\
reproduction with mutation only) or help avoid deadlocks (when the number of
agents in the environment is lower than the capacity of the arena).

Listing~\ref{ls:asyncFight} shows an example of a meeting in the fighting arena.
As the arena has no direct access to agents state it needs to query them with the
use of messages. A Scala feature known as \emph{futures} and \emph{for
comprehension} allows to implement asynchronous and non-blocking meetings. The
\texttt{askForFitness} and \texttt{getEnergyFrom} functions return a
\emph{future value} which will be \emph{completed} only when all the members
reply to messages. \emph{For comprehension} composes these futures into a new one which
is returned from the \texttt{performMeeting} function, allowing installation of a 
\emph{completion hook} (Listing~\ref{ls:asyncArena}, lines 6 to 9).
The important thing is that the \texttt{performMeeting} function can return before the
meeting has actually ended, so that another meeting may be triggered in the
arena.

\begin{figure}[h!]
\begin{lstlisting}[language=Scala, label=ls:asyncFight,
caption={Non-blocking asynchronous fight using Scala futures and
\emph{for comprehension}.}] 
def performMeeting(members) = for( 
		fitnesses <- askForFitness(members);
		val winner = zip(members, fitnesses)
			.maxBy { (m, f) => f }
			.map { (m, f) => m }
		val losers = members - winner;
		energies <- getEnergyFrom(losers)
		) yield winner tell ReceiveEnergy(energies.sum)
\end{lstlisting}
\end{figure}

\subsection{Synchronous EMAS}
\label{sec:syncEmas}

In this version, agents are considered parts of the model rather than the
implementation, like in Netlogo. As such, they are not represented as individual
entities but as data structures.

Populations are collections of agents, processed step by step by arenas to yield
new collections (see Listing~\ref{ls:syncAgents}). Agents are split into arenas
(lines 2-4) and grouped accordingly to the arity of each arena (line 7).
Finally, groups are processed by arenas and the agents resulting from each
meeting are combined into a new population (lines 5-10).

\begin{figure}[h!]
\begin{lstlisting}[language=Scala, label=ls:syncAgents,
caption={Agents are split between arenas, grouped and processed.
These action repeatedly transform the population.}] 
def step(population) = { 
	val agentsInArenas = population groupBy { agent =>
		agent.chooseArena 
	}
	val newPopulation =  agentsInArenas flatMap { 
		(arena, agents) => 
			agents grouped(arena.size) flatMap { 
				members => arena.performMeeting(members)
			}
	}
	return newPopulation shuffled
\end{lstlisting}
\end{figure}

The \texttt{performMeeting} method of each arena should in this case return a
collection of agents representing the result of a meeting. These collections are
merged into the new population. The Listing~\ref{ls:syncFight} shows the
implementations of a synchronous fighting arena, which is similar but simpler
than in the asynchronous version, as arenas can now have direct and synchronous
access to the state of agents.

\begin{figure}[h!]
\begin{lstlisting}[language=Scala, label=ls:syncFight,
caption={A synchronous fighting arena transforms its members by transferring
energy from losers to winners.}] 
def performMeeting(members) = { 
	val winner = members maxBy { 
		agent => agent.fitness 
	}
	val losers = members - winner
	val energies = getEnergyFrom(losers)
	winner.energy += energies.sum
	return members
}
\end{lstlisting}
\end{figure}

The step function from Listing~\ref{ls:syncAgents} could be executed in a simple loop.
However, we used an Akka actor which repeatedly sends a Step message to itself,
in order to minimise the performance impact of the Akka framework itself when
comparing both versions.

It should be added that the structure of the synchronous version is similar to
the MapReduce pattern and could be parallelised in a similar way. While this is
a topic of the current research, we decided to stick to a possibly simple version in
this work.

%% file: experiments.tex
\section{Experimental Results}
\label{sec:experiments}

\begin{figure}[!h]
	\centering
	\begin{minipage}{.45\textwidth}
        \centering
        \includegraphics[width=\textwidth]{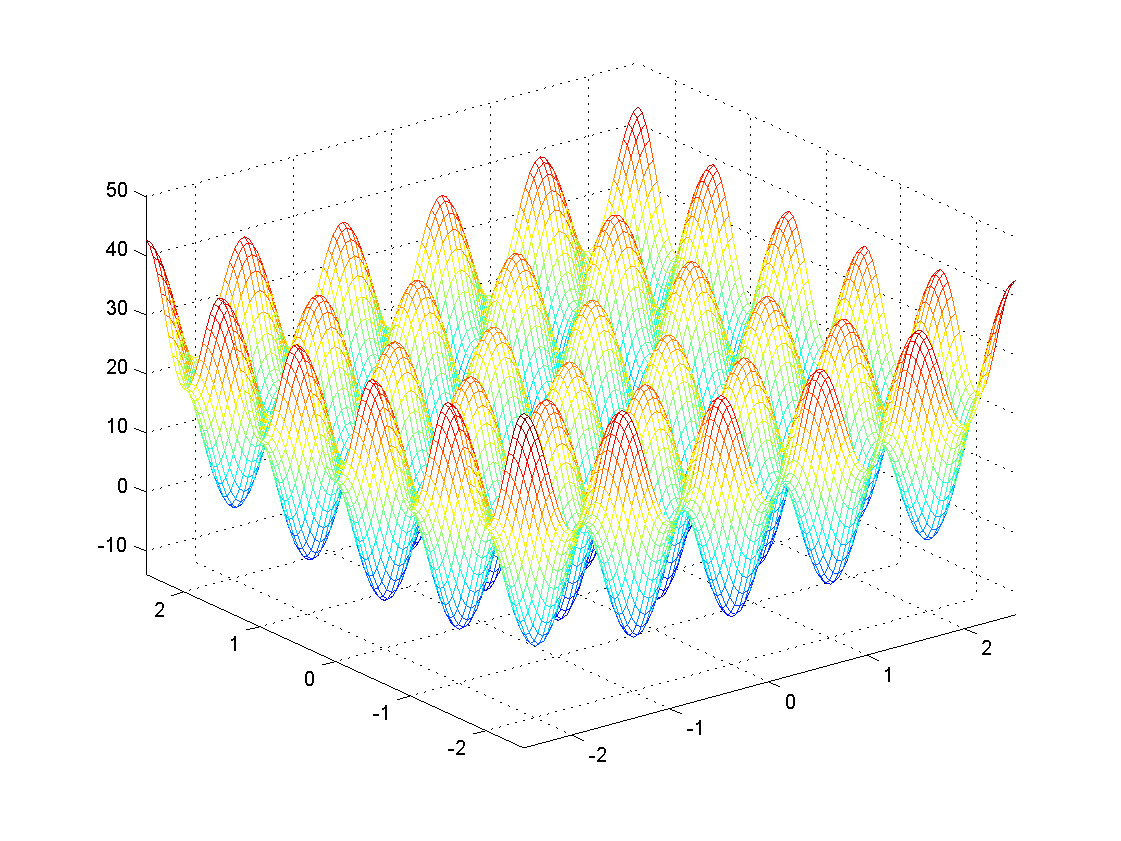}
    \end{minipage}
    ~
    \begin{minipage}{.45\textwidth}
        \centering 
        \includegraphics[width=\textwidth]{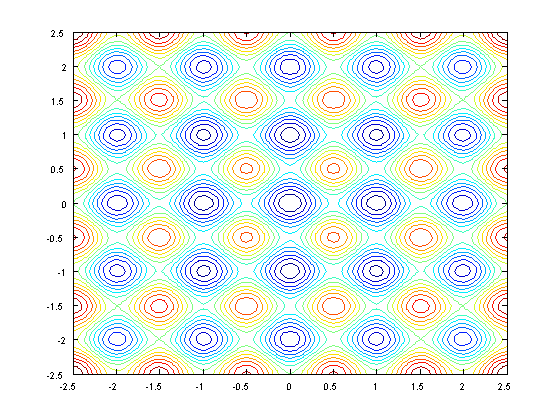}
    \end{minipage}
    \caption{The Rastrigin function in two dimensions.}
    \label{fig:rastrigin}
\end{figure}

We carried out a series of experiments to measure the performance and
scalability of the implementations described in the previous section.
We applied the evolutionary multi-agent system to the optimisation task of
finding the global minimum of the Rastrigin benchmark function
(Eq.~\ref{eq:rastrigin}), a highly multimodal function with many local optima
and one global minimum equal 0 at $\bar{x} = 0$ (Fig.~\ref{fig:rastrigin}). We
used a problem size (a dimension of the function) equal to 100.

\begin{equation}
f(x) = 10n + \sum\limits_{i=1}^n(x^2_i - 10 \cos(2 \pi x_i))
\label{eq:rastrigin}
\end{equation}

The parameters used in our experiments are listed in Table~\ref{tab:emas}.

\begin{table}[!h] \centering
\begin{tabular}{|l|l|}
  \hline 
	initial-size & 50 \\
	initial-energy & 10 \\
	reproduction-threshold & 10 \\
	reproduction-transfer & 5 \\
	fight-transfer & 10 \\
	fight-arena-size & 2 \\
	migration-probability & 0.001 \\
  \hline
\end{tabular} 
~~~~~~~
\begin{tabular}{|l|l|}
  \hline 
	problem-size & 100 \\
	mutation-rate & 0.1 \\
	mutation-range & 0.05 \\
	mutation-probability & 0.75 \\
	recombination-probability & 0.3 \\
	 &  \\
	 &  \\
  \hline
\end{tabular} 
\caption{EMAS parameters.}
	\label{tab:emas}
\end{table}

The environment was initialised with \emph{initial-size} agents, each given
\emph{initial-energy}. Agents were fighting on an arena of
\emph{fight-arena-size} capacity, transferring \emph{fight-transfer} energy from
a loser to a winner. As soon as an agent's energy exceeded
\emph{reproduction-threshold}, it entered a reproduction arena of size 2.
Each pair of agents in the reproduction arena reproduced using a set of genetic
operators described below, creating 2 new agents, each one given
\emph{reproduction-transfer} energy from one of their parent.

In the second stage of our experiments, at each step every agent had a
\emph{migration-probability} of migrating to some other environment. The target
environment was chosen at random and including the original one.

In order to create new solutions to be assigned to newborn agents, the following
genetic operators were used. Solutions were encoded as real-valued vectors. At
each reproduction, crossover and mutation happened with respectively
\emph{recombination-probability} and \emph{mutation-probability}. We used random
average crossover, which consist in picking a random point in the hypercube
defined by the parents genotypes. Every feature in the solution vector was
mutated with probability \emph{mutation-rate}. We used gaussian mutation with
\emph{mutation-range} standard deviation.

\subsection{Performance Testing}

In our performance testing, we distinguished four experimental scenarios. All of
them share the same set of parameters listed above and have been run on Pl-Grid
\footnote{\url{http://www.plgrid.pl/en}} infrastructure. We used nodes with an Intel
Xeon X5650 2,66 GHz processor, with 1GB of memory and a variable number of
active cores (up to 12).

The first three scenarios correspond to the asynchronous implementation with an
Akka dispatcher configured with respectively \emph{own-thread},
\emph{thread-pool} and \emph{single-thread} policy (see Section~\ref{sec:asyncEmas}).
The fourth scenario is the synchronous implementation
(Section~\ref{sec:syncEmas}).

Each scenario was repeated 30 times with different random generator seed
values. The asynchronous models have been run for 60 minutes each, while the
synchronous one only for 10 minutes. 

We gathered two metrics: 
\begin{inparaenum}[a)]
	\item the fitness of the best solution found so far at any given time,
	\item the number of fitness function evaluations at any given time.
\end{inparaenum}
The former metric shows the efficiency of the evolutionary algorithm itself and is also dependent on i.e. the parameters of the evolutionary operators. The latter reflects the number of agent meetings and only depends on the execution model and threading strategy. Of course, the dynamics of the underlying multi-agent system have an impact on the efficiency of the evolutionary algorithm.

\begin{table}[!h] 
\centering
\begin{tabularx}{0.9\textwidth}{X|X|X|X|X|X|X}
  	\cline{2-6} 
  	 & \multicolumn{5}{c|}{cores} &\\
  	\cline{2-6}
    & 1 & 2 & 4 & 8 & 12 &\\
    \cline{1-6}
	\multicolumn{1}{ |l| }{own} & 22.1500 & 19.4800 & 12.7577 &  9.2573 &  3.6087
	&\\
	\multicolumn{1}{ |l| }{pool} & 18.0173 & 17.6303 & 11.1574 &  5.4219 & 
	3.8016 &\\
	\multicolumn{1}{ |l| }{single} & 15.2845 & 19.2584 &  6.3323 &  8.1770 & 
	3.4879 &\\
	\cline{1-6}\cline{1-6}
	\multicolumn{1}{ |l| }{sync} & 0.0371 & 0.0398 & 0.0321 & 0.0186 & 0.0257 &\\
  	\cline{1-6}
\end{tabularx}  
 \caption{Final best fitness found in each of the models with
 a given number of cores. The asynchronous models were run 60 minutes, the
 synchronous one was run 10 minutes. The results are averaged over 30 runs.}
 \label{tab:fitsTable}
\end{table}

\begin{table}[!h] 
\centering
\begin{tabularx}{0.9\textwidth}{X|X|X|X|X|X|X}
  	\cline{2-6} 
  	 & \multicolumn{5}{c|}{cores} &\\
  	\cline{2-6}
    & 1 & 2 & 4 & 8 & 12 &\\
    \cline{1-7}
	\multicolumn{1}{ |l| }{own} & 1.0407 & 1.5252 & 1.6836 & 2.0900 & 2.9918 &
	\multicolumn{1}{ l| }{\multirow{4}{*}{$\times 10^7$}}\\
	\multicolumn{1}{ |l| }{pool} & 1.4270 & 1.7994 & 2.0061 & 2.6123 & 2.8876 &
	\multicolumn{1}{ l| }{}\\
	\multicolumn{1}{ |l| }{single} & 1.7423 & 1.4949 & 2.2076 & 2.2561 & 2.9611 &
	\multicolumn{1}{ l| }{}\\
	\cline{1-6} \cline{1-6}
	\multicolumn{1}{ |l| }{sync} & 3.3296 & 3.1524 & 3.7204 & 5.6854 & 4.6629 &
	\multicolumn{1}{ l| }{}\\
  	\cline{1-7}
\end{tabularx} 
 \caption{Total number of fitness evaluations in each of the models with
 a given number of cores. The asynchronous models were run 60 minutes, the
 synchronous one was run 10 minutes. The results are averaged over 30 runs.}
 \label{tab:evalsTable}
\end{table}

 The results in Tables~\ref{tab:fitsTable} and \ref{tab:evalsTable} indicate
that:
\begin{compactitem}
	\item There was no statistically significant difference in the performance
	of the asynchronous version using different thread policies (as verified by a two-sample Kolmogorov-Smirnov test with p=0.5).
	\item The asynchronous versions greatly improved when given more
		cores\ldots
	\item \ldots but were \emph{dramatically} worse than the synchronous version.
\end{compactitem}

\begin{figure}[!h]
	\centering
    \includegraphics[width=.7\textwidth]{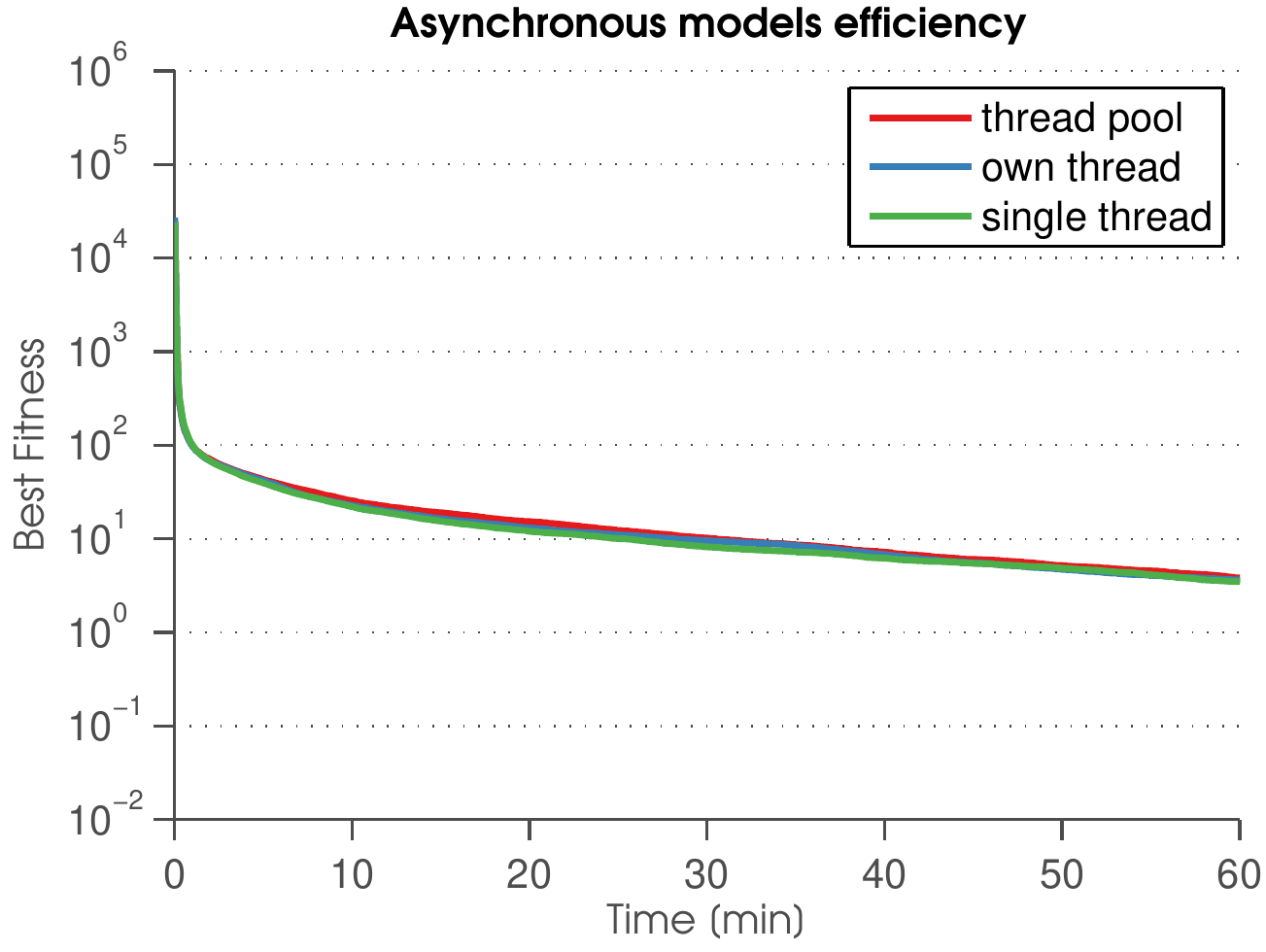}
    
	\includegraphics[width=.7\textwidth]{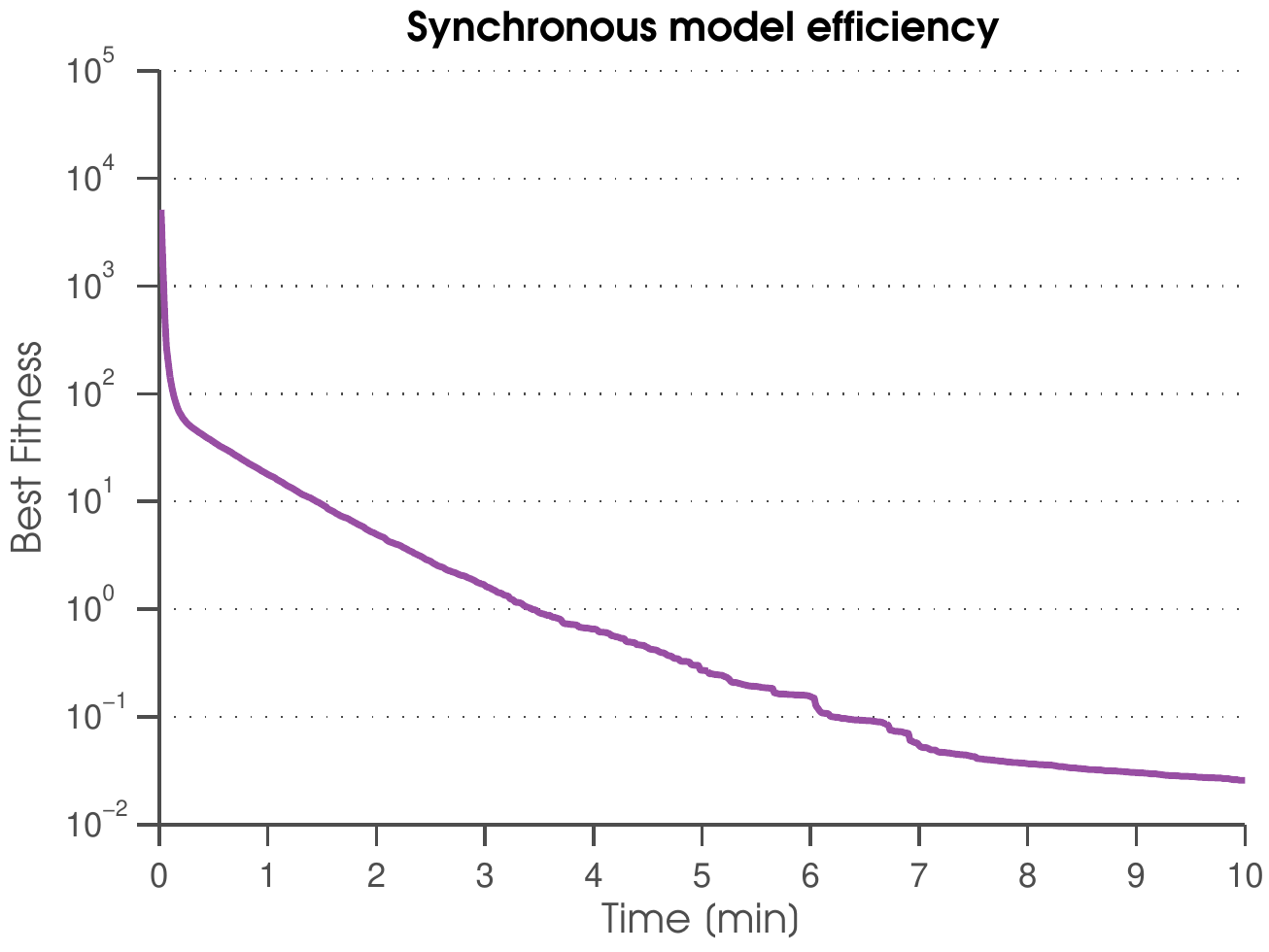}
    
	\caption{Average best fitness found in each of the models after a given
    amount of time (for 12 cores used). Top -- asynchronous models, bottom
	-- synchronous model.}
    \label{fig:bestFitness} 
\end{figure}

\begin{figure}[!h]
	\centering
    \includegraphics[width=.7\textwidth]{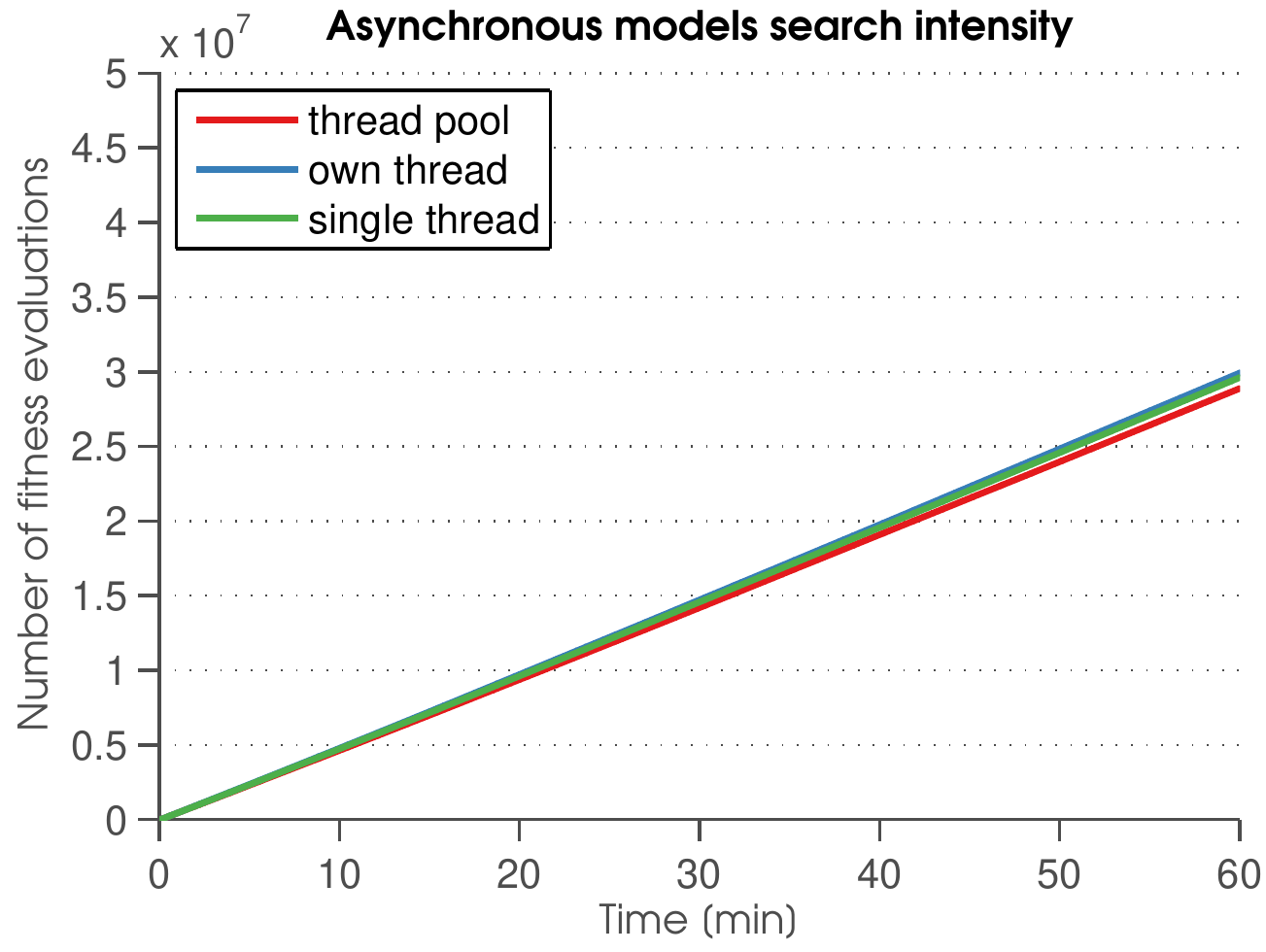}
    
	\includegraphics[width=.7\textwidth]{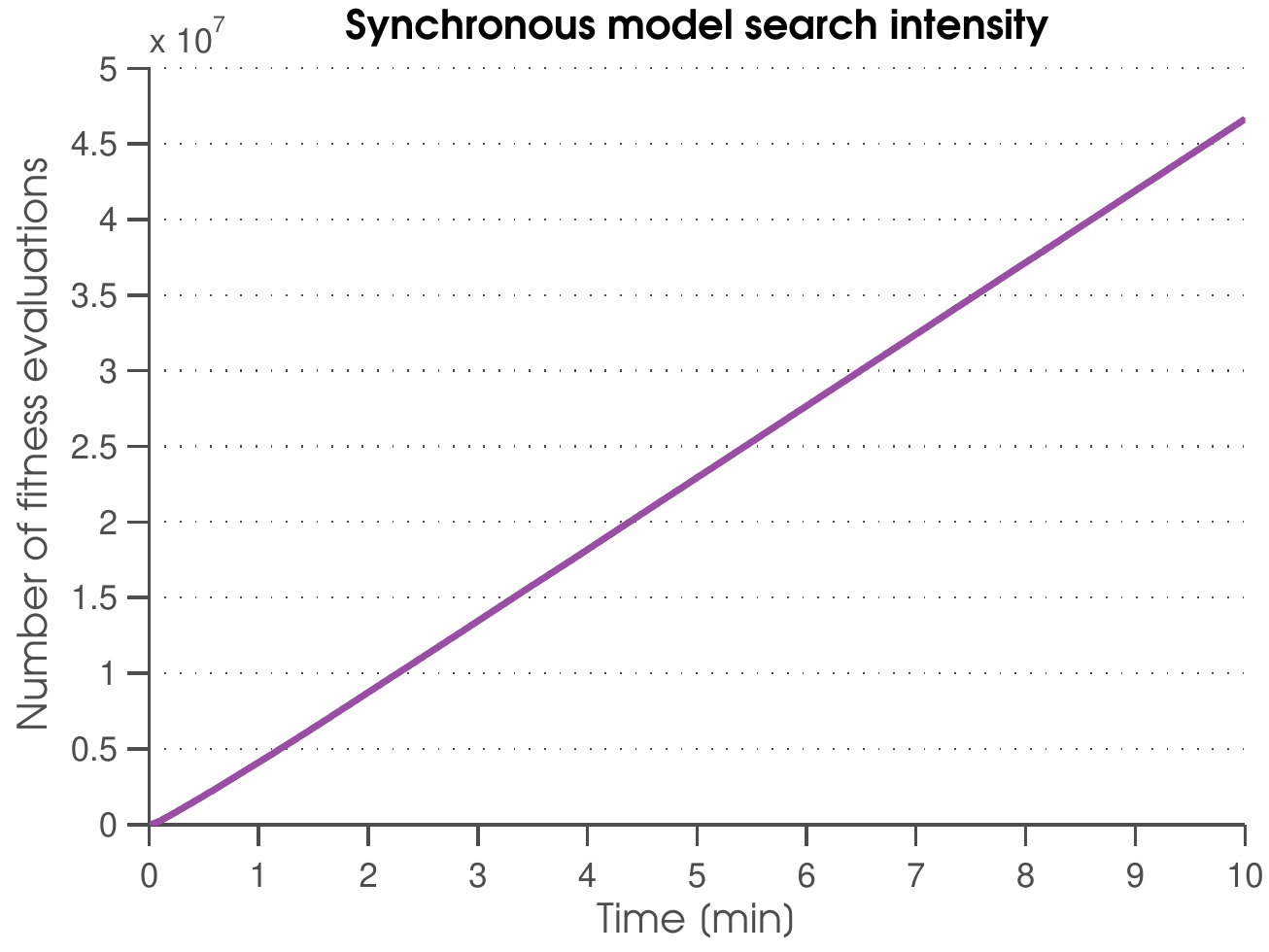}

    \caption{Amount of fitness evaluations performed in each of the models after
    a given amount of time (for 12 cores used). Top -- asynchronous models, bottom
	-- synchronous model.}
    \label{fig:evalsCount}
\end{figure}

This difference in efficiency did not come from a flaw in the evolutionary
algorithm itself, but rather from the underlying implementation model. The best
asynchronous version only performed about $8 \times 10^3$ fitness evaluations
per second, while the synchronous version did more than $7 \times 10^4$ --- nearly
an order of magnitude faster. This efficiency gap can clearly be seen in
Figures~\ref{fig:bestFitness} and \ref{fig:evalsCount}. 

Profiling data suggest that the asynchronous implementation was not idle or blocking
on I/O, but rather very busy managing threads and passing messages between
actors.

\begin{figure}[!h]
	\centering    
	\includegraphics[width=.7\textwidth]{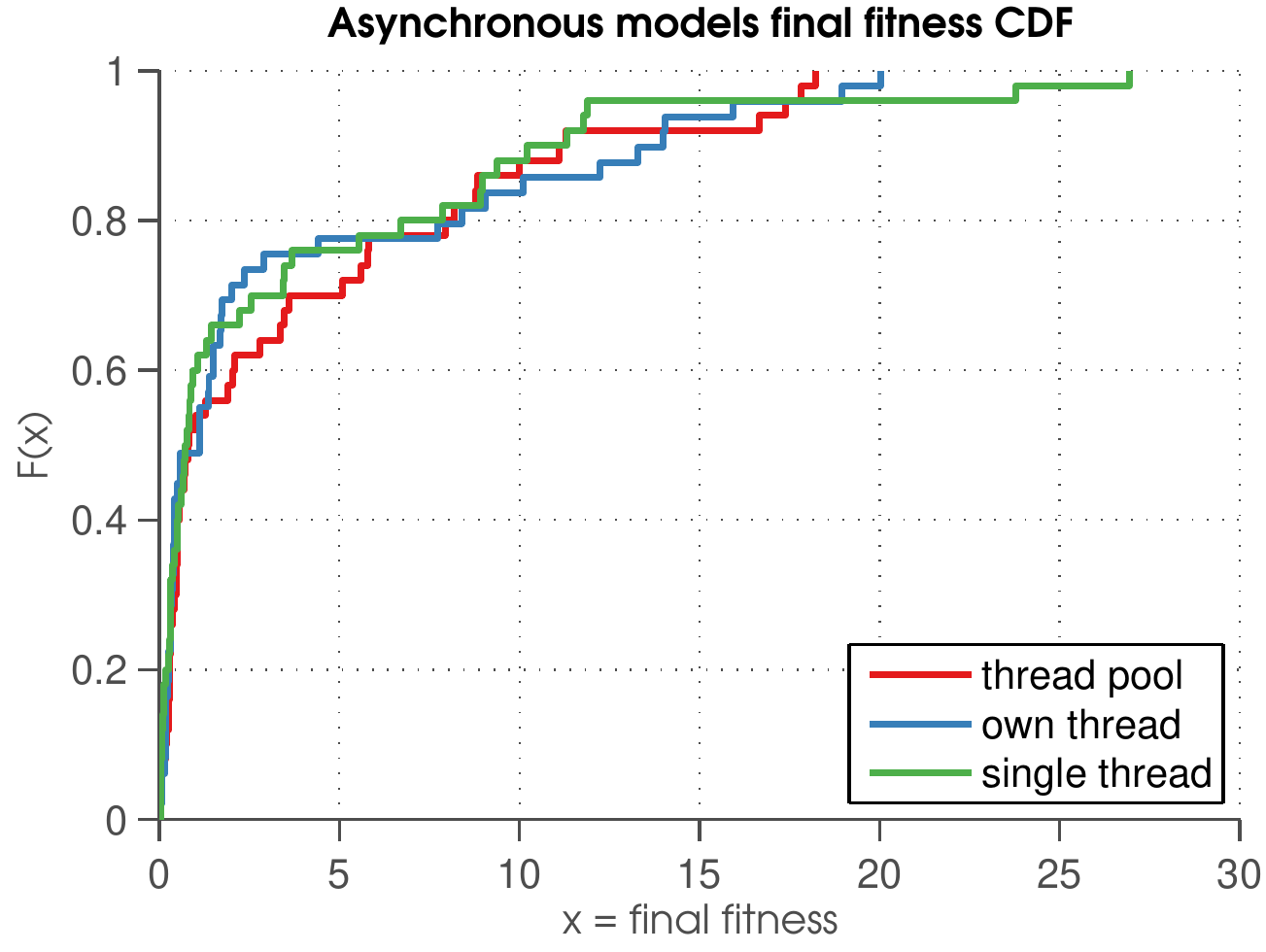}
       
	\includegraphics[width=.7\textwidth]{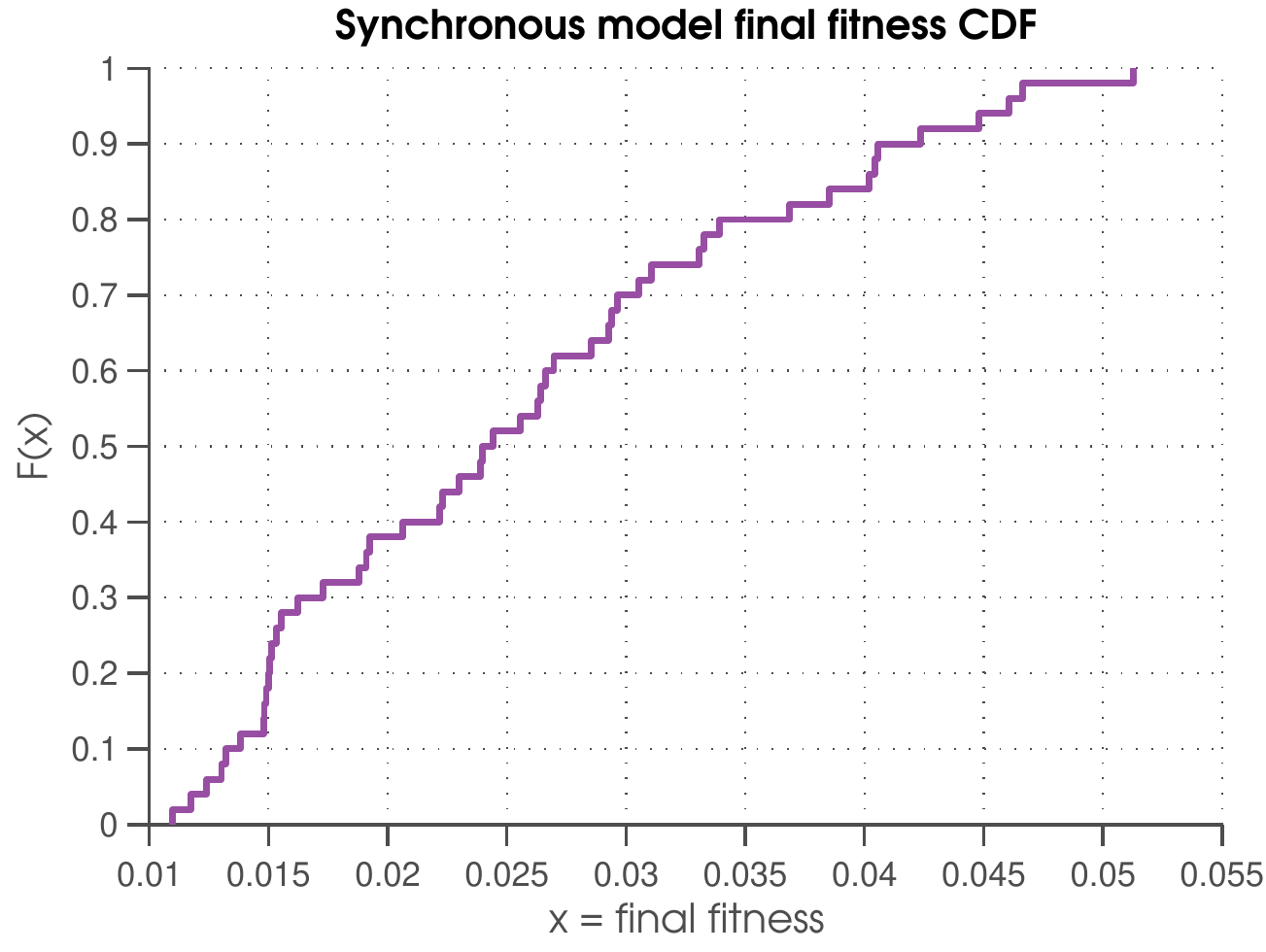}

    \caption{Empirical cumulative distribution functions of the final
    fitness values found at the end of the computations (for 12 cores used).
	Top -- asynchronous models, bottom -- synchronous model.}
    \label{fig:bestFitnessCdf} 
\end{figure}

\begin{figure}[!h]
	\centering
    \includegraphics[width=.7\textwidth]{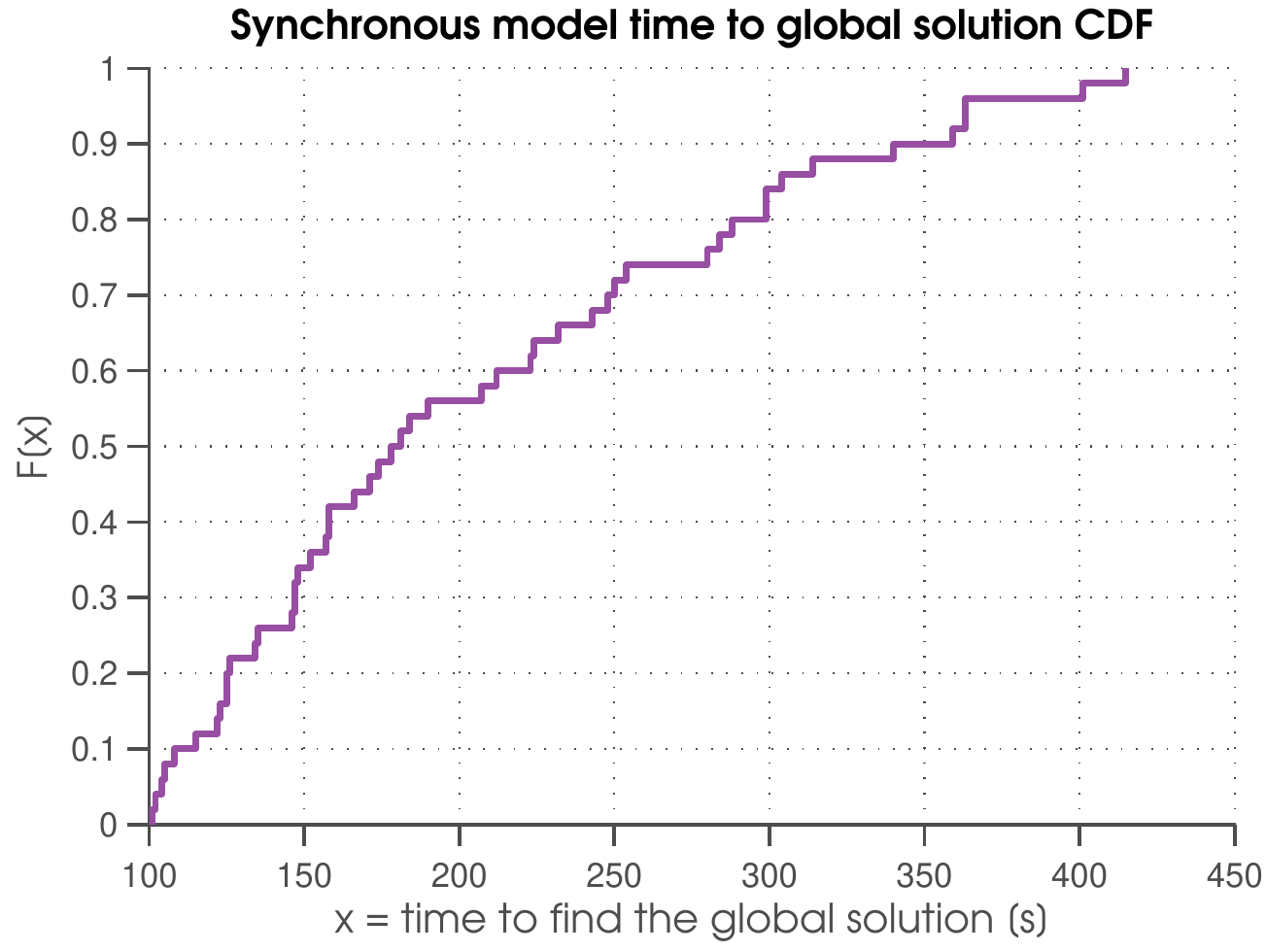}
    \caption{Empirical cumulative distribution function of the time required to
    find the global solution in the synchronous model (for 12 cores used).}
    \label{fig:winTime}
\end{figure}

Figure~\ref{fig:bestFitnessCdf} shows the empirical distribution functions
of the final fitness achieved in separate runs of each model. In many cases, the
asynchronous version did converge to acceptable solutions (though given much
more time). However, in many runs, they clearly needed more time. In contrast,
all the runs of the synchronous version converged to the attraction basin of the
global optimum (which corresponds in the case of the Rastrigin function to a
fitness value lower than 1). It took an average of 132.13 s ($\pm$ 23.82 s), the
empirical cumulative distribution is shown in Figure~\ref{fig:winTime}.

\subsection{Scalability Testing}

Having determined that the synchronous model is more efficient, we went on to
test the scalability of the algorithm when new resources were added. We modified
the implementation to simultaneously start a synchronous EMAS environment on
many nodes in a cluster. The environments discovered each other in the cluster
and connected, enabling the migration of agents between them.

We considered several scenarios with an increasing number of nodes. Each
scenario was run for 10 minutes and repeated 30 times.

\begin{figure}[!h]
	\centering
	\includegraphics[width=.7\textwidth]{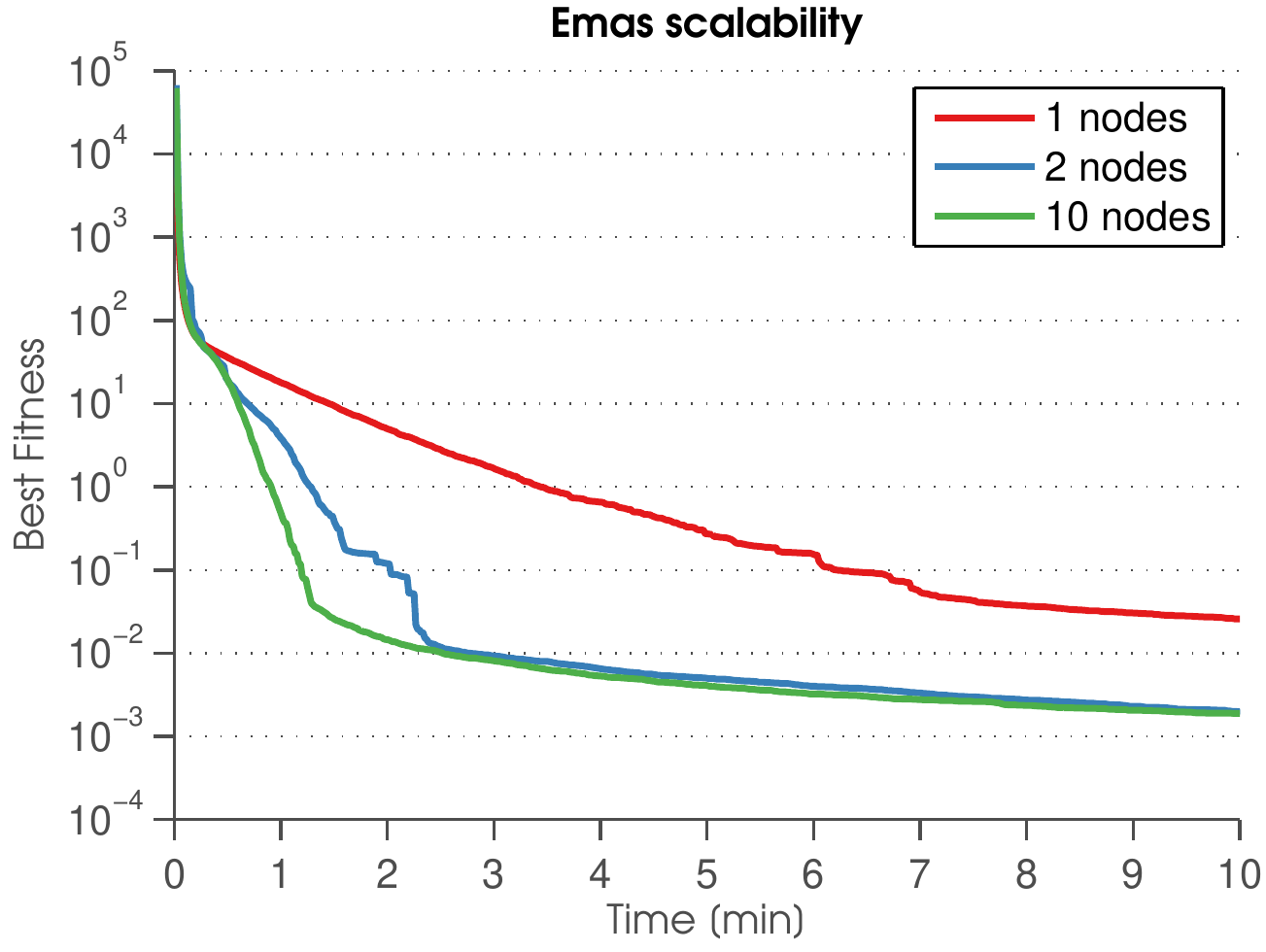}
    \caption{Best fitness reached within a given time when using different
    amounts of nodes (parallel environments).}
    \label{fig:clusterFitness}
\end{figure}

Fig.~\ref{fig:clusterFitness} shows the fitness of best solution found after
a given time, averaged over all environments in a given scenario and all runs
of the scenario. We can see that even adding a second node to the
computation leads to significantly better results. Moreover, adding more nodes
increases the convergence rate.

These results show that is efficient to decompose multi-agent systems into
distributed environments, like in the classical island model of evolutionary
algorithms. However, the decentralised semantics of agent interaction may lead
to more intelligent migration strategies, for example where agent populations
automatically balance the load in the cluster.